\documentclass{article}

\usepackage{amsmath, amsthm, amsfonts}
\usepackage{graphicx}  
\usepackage{amsmath}
\usepackage{amssymb}
\usepackage{siunitx}
\usepackage{array}
\usepackage{color}
\usepackage[utf8]{inputenc}

\title{Macroscopic quantum tunneling in PdO nanoparticles}
\author{Francisco Ascencio  \and C. Reyes-Damián   \and Roberto Escudero\\
  \small Instituto de Investigaciones en Materiales, Universidad Nacional Aut\'onoma de M\'exico\\
  \small Mexico City, 04510 Mexico\\
  \small fascencioag@gmail.com
}

\begin{document}
\maketitle

\abstract{We studied the physical behavior of  PdO nanoparticles at low temperatures, which presents an unusual behavior clearly related to macroscopic quantum tunneling. 
The samples show a tetragonal single phase with P42/mmc space group. Most importantly, the particle size was estimated at about 5.07 $\pm$ 1.97 nm. Appropriate techniques were used to determine the characteristic of these nanoparticles. 
The most important aspect of this study is the magnetic characterization performed at low temperatures. It shows a peak at 50 K in zero field cooling mode (ZFC) that corresponds to the Blocking temperature ($T_{B}$). These measurements in ZFC and field cooling (FC) indicate that the peak behavior is due to different relaxation times of the asymmetrical barriers when the electron changes from a metastable state to another. Below $T_{B}$ in FC mode, the magnetization decreases with temperature until 36 K; this temperature is the crossover temperature $T_{Cr}$ related to the anisotropy of the barriers, indicative of macroscopic quantum tunneling.
}

\section{Introduction}
\label{intro}
The study of the magnetic properties of materials has been a topic of great interest in nanoparticles because of the unusual new properties. For instance, molecular magnets offer interesting characteristics useful to different systems. In general, compounds formed by nanostructures could be one of the most simple manners to observe the effects of macroscopic quantum tunneling , MQT.

One relevant characteristic seen in ferromagnetic materials is that their magnetic characteristics change when the size is reduced to a limit; this is the critical value. In this limit, the magnetic behavior changes to superparamagnetism, SP. At that point, the thermal assisted SP can be blocked by random thermal fluctuations due to spin orientations and could be frozen as the temperature decreases. At low temperatures, the thermal fluctuations are frozen and consequently become trapped and can not surpass the magnetic barriers of metastable states. At this point, quantum tunneling magnetic effects,  MQT takes place because nanoparticles show a dramatic increase in the surfaced to volume ratio; the effect occurs because of anisotropy of the barriers. At this point, MQT  clearly is observed, and also other characteristics as spin glass, behavior, paramagnetic, and or superparamagnetism characteristics.

Experimentally some of those features related to the MQT can be seen at low temperatures:  the blocking and crossover temperatures. Both features are quite dependent and related to the anisotropy of the arising energy barriers in the metastable states that allow the spin transference between two states. 

The characteristics and properties observed in nanomaterials are intrinsic properties of electronic interaction between them. In 3d and 4d series atoms, band structure changes dramatically, changing the surface because of the narrowing of the bands and distances. One aspect that is important in the study of the cumulous of nanoparticles. Accordingly, to carry out and understand this behavior is quite important to know the coordination of the structures formed by the cumulous.

Pd samples with macroscopic dimensions behave quite differently as nanosize scale; this behavior was observed by many authors \cite{oba2008ferromagnetism,shinohara2003surface,xiao2009shape,jeon2008magnetism}. 
  
When nonmagnetic material is in the critical size \cite{cox1994magnetism,nealon2012magnetism} studies performed in Metal-Oxygen compounds show that the magnetic behavior is strongly related to oxygen vacancies, substitutional and interstitial defects, as well as surface defects of the nanomaterials.\cite{gao2010vacancy,punia2021oxygen,choudhury2013room} However, one of the most unusual magnetic characteristics in  nanomaterials is the observation of   properties that can be related to macroscopic quantum tunneling of the magnetization (QTMs or MQT) \cite{chudnovsky1988quantum,chudnovsky1993macroscopic,barbara1995mesoscopic,zhang1996magnetic,kodama1999magnetic,thomas1998quantum,barbara1999magnets,wernsdorfer1998magnetization,barbara1997quantum}. 
The MQT has been observed mainly in nanoparticles compounds of metal oxides; the two important characteristics or features can be observed at low temperatures, they are the blocking $T_B$  and the cross-over temperatures, $T_{Cr}$.\cite{zhang1994time,kodama1994low,zhang1995magnetic,ibrahim1995nonlinear,xiao2012quantum}

It is important to mention that the study of metallic oxides is attractive because of numerous applications  that can be   used in different fields; i.e.  catalysis, as precursors to the synthesis of superconductors, electronic devices, biomedical applications, and gas sensors \cite{vedrine2019metal,cava2000oxide,greiner2010metallic,shahbazi2020versatile,nunes2019metal}. Additionally  these studies become relevants when the compounds are  low-dimensional materials as in thin films, wires, or nanoparticles. Nowadays, it is well known that size, shape, and surface structure, and physicochemical properties as in optical, electronics, and magnetism  behaves differently than in bulk. Related to this study,  Palladium Oxide, PdO,  nanoparticles have attracted attention because of the uniques physicochemical properties. PdO in bulk is a nonmagnetic compound \cite{ahuja1994optical}. It presents a tetragonal symmetry, Cooperite, in which Pd atoms occupy D$_{2h}$ and O atoms occupy D$_{2d}$ positions \cite{hass1992band}. In the structure, the Pd atoms are coordinated by 4 oxygen atoms in a planar manner, while  O atoms are coordinated with 4 Pd atoms, forming a tetrahedron. PdO is a P-type semiconductor, with a direct gap. In previous studies by many authors has been observed that the bandgap is close to 2 eV \cite{kumari2019sol,bardhan2013size} which has industrial interest mainly because its photocatalytic characteristics \cite{veziroglu2020pdo} used in the oxidation of CH$_{4}$ \cite{mccarty1995kinetics}, and in sensors \cite{das2014highly}.
Actually , some methods have been used to synthesize PdO nanostructures; as sol-gel\cite{zhao2019study},thermo-hydrolyzation assisted by microwave \cite{li2020mono}. Related to  this work the synthesize was performed using a simple method that a continuation will be described.

  \section{Experimental Details}
\label{Exp}

PdO nanoparticles were synthesized by an alkali salt method, synthesized with concentrations of 0.125 g of  Pd(NO$_{3}$)$_{2}$ * 2 H$_{2}$O (Sigma Aldrich, 99 \%), 0.125 g of Li$_{2}$CO$_{3}$ (Ripley Scientific) and 2.5 g of NaCl (Sigma Aldrich)  were mixed in an agate mortar and milled for 30 minutes until obtain a fine powder. Samples were heated at 400 $^{\circ}$C in a furnace at atmospheric pressure. Finally, the obtained samples were washed five times with distilled water in order to eliminate residues of the reaction.

\subsection{Characterization}
\label{Chrc}
Characterized of the samples were carried out by X$-$ray diffraction (XRD) performed with a Bruker (D8 Advance) and  Cu$-$K$_{\alpha}$ radiation, determined from  2$\Theta$ range of 20$^{\circ}-120^{\circ}$. The phase was identified using the Crystallographic Open Database (COD) and crystalline structure refined with Rietveld method using the BGMN program \cite{bergmann1998iucr} and the graphical interface Profex. Peak profiles were modeled to determine the crystal size using the Debye$–$Scherrer formula \cite{klug1974x}. Transmission electron microscopy (TEM), JEOL JEM-2010F FasTem microscope operated at 200 kV. The samples for  TEM  were prepared, depositing a drop of the solution into the mesh Cu grid. The Raman spectra were recorded in a Thermo Scientific confocal microscope equipped with micro$-$Raman using three different excitation sources: 532 nm (2.33 eV), 633 nm (1.95 eV), and 780nm (1.58 eV); measurements were carried out in the 50$–$3000 cm$^{-1}$ range. Optical properties were characterized by UV$–$visible spectroscopy with a Thermo Evolution 220 spectrophotometer. Finally, magnetization as a function of temperature (M(T)) and magnetization as a function of the magnetic field (M(H)) were determined using a Quantum Design magnetometer (MPMS).

\section{Results and Discussion}
\label{Res} 
\subsection{X-Ray Diffraction}
\label{X-Ray}

\begin{figure}[h!] 
\centering
\includegraphics[width=0.7\textwidth]{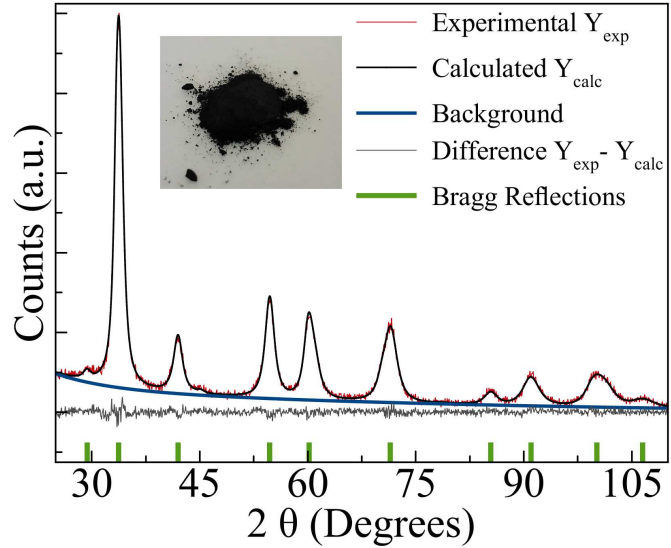} 
\caption{X-ray diffraction pattern measured with CuK$_{\alpha}$ radiation and bragg reflections of reference $96-100-9032$ taken from the Crystallographic Open Database. The right inset shows the Rietveld refinement of PdO nanoparticles diffraction pattern. The bottom line is the difference between experimental and calculated pattern. A photograph of the powder sample is also shown in the left inset.}
\label{Fig:1}        
\end{figure}  

The crystallographic structure of the PdO nanoparticles was determined by X-ray diffraction (XRD). XRD pattern is shown in  the figure  \ref{Fig:1}, there we show the XDR measurements of PdO (nano)powders in which the PdO tetragonal phase was detected. The right inset of the figure \ref{Fig:1} shows the Rietveld refinement of the sample; the red line is the experimental data, in the inset, the black line is the calculated pattern. At the bottom of the diagram, the differences between the experimental and calculated points are shown in gray color. Vertical blue marks at the bottom are displayed and correspond to the Bragg positions for the phase. Finally, the background was modeled with a Lagrangian polynomial. From Rietveld refinement, we see the phase identified and modeled with tetragonal PdO phase according to the Crystallographic Open Database (COD) number $96-100-9032$ described by the space group P42/mmc. The average crystal size was 10.31 $\pm$ 0.83 nm. Cell parameters were calculated the resulting lattice parameters were a= 0.3042 $\pm$ 0.0001 and c = 0.5356 $\pm$ 0.0003 nm, according to the space group \# 131. Finally, Rietveld residues were: Rwp=6.91\%, Rexp=6.51\% and Rwp/Rexp = 1.06.

\subsection{Transmission Electron Microscopy}
\label{E. M.}

\begin{figure*}[htb]
\centering
\includegraphics[width=0.5\textwidth]{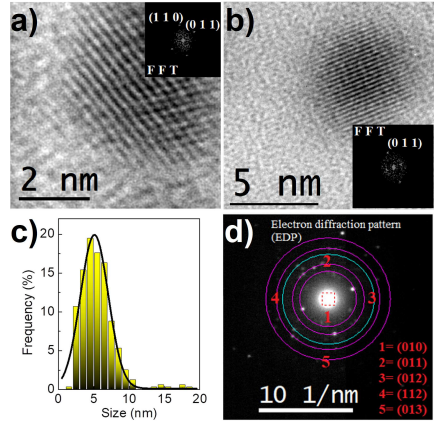}
\caption{\textbf{(a)}HRTEM image of a very small PdO nanoparticles with the corresponding FFT, \textbf{(b)} another HRTEM image for a representative nanoparticle with the corresponding FFT,\textbf{(c)} the nanoparticles size distribution \textbf{(d)} electron diffraction pattern.}
\label{Fig:2}
\end{figure*}

Transmission electron microscopy was carried out to obtain structural and morphological information of the samples. In the figure \ref{Fig:2} (a), HRTEM micrograph is presented. It is observed that the nanoparticles have an irregular polyhedron shape. The first particle of around 5 nm presents the interplanar distances d$_{1}$ = 2.195 \r{A} and d$_{2}$ = 2.523 \r{A}, with an angle between planes of $\theta$= 52.07$^{\circ}$. These reflections correspond to the (1 1 0) and (0 1 1) planes oriented along the [-1 1 -1] axis zone from the FFT lattice fringes are resolved. Figure \ref{Fig:2} (b) shows another nanoparticle with 4 nm of diameter with the FFT, is seen only the plane (0 1 1). In the figure \ref{Fig:2} (c), the nanoparticles average diameter were  5.07$\pm$1.97 nm. measured using a total of 200 particles. The structure of these nanoparticles was confirmed through electron diffraction, figure \ref{Fig:2} (d), the pattern was matched with the PdO crystal structure. The reflections were indexed as the (0 1 0), (0 1 1), (0 1 2), (1 1 2), and (0 1 3) crystallographic planes, according to COD, \# $96-100-9032$. 
The optical and magnetic properties of palladium oxide nanoparticles are related to grain boundaries or defects in the nanoparticle structure; hence, the HRTEM images permit a clear observation of their fine features. With TEM studies, it can be observed that the produced nanoparticles are single crystals without the presence of twins or grain boundaries.

\subsection{Raman and UV-visible Spectroscopy  }
\label{R. S.}

Raman spectra were collected at room temperature from the sample powders to obtain detailed structural information related to normal vibrational modes of the samples with three wavelengths $\lambda$= 532, 633, and 780 nm or excitation energies of 2.33, 1.95, and 1.59 eV, in order to observe resonance phenomena.

Assuming that the PdO compound has a tetragonal symmetry with space group P42/mmc (\#131) and point group D4$_{h}$, and according to the factor group analysis, this compound will display two vibrational Raman active modes (B$_{1g}$ and E$_{g}$). The remaining normal modes of vibration are located in the IR region (A$_{2u}$, 2E$_{u}$) and a silence mode B$_{2u}$ \cite{remillard1992optical,bardhan2013size}. 

\begin{figure}[h!] 
\centering
\includegraphics[width=0.5\textwidth]{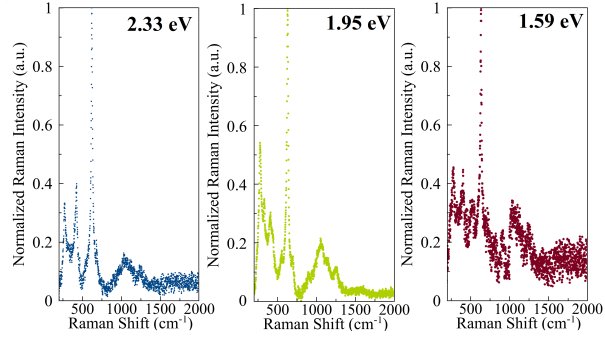} 
\caption{Full Raman spectra of PdO nanoparticles at room temperature for three excitation energies 2.33, 1.95 and  1.59 eV}
\label{Fig:3}       
\end{figure}

In the figure \ref{Fig:3}, the normalized Raman spectra of the samples measured at room temperature with  three different wavelengths (excitation energy) are shown. It is observed that the spectra show some Raman peaks below 1500 cm$^{-1}$, in Table \ref{tab:1} a list of the peaks are presented for the three excitation energies.

\begin{figure}[h!] 
\centering
\includegraphics[width=0.5\textwidth]{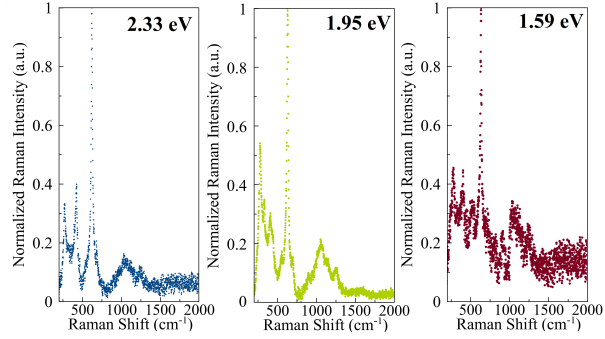} 
\caption{Comparison of Raman spectra  of PdO nanoparticles around the main peak  for different excitation energy} 
\label{Fig:4}       
\end{figure}

A careful analysis of the Raman spectra (Figure \ref{Fig:4}) for each  energy shows that the PdO nanoparticles have  three main peaks related to B$_{1g}$  between the frequencies: [620.62$-$630.83] cm$^{-1}$, peaks between the frecuencies [425.09$-$401.66] cm$^{-1}$ which corresponds to E$_{g}$ and finally the vibrational mode between [273.01$-$278.24 ] cm$^{-1}$ correspond to X$_{8}$ mode \cite{zhao2019study}. Besides many other peaks related to second order processes overtones, prohibited modes and combinations of normal modes were observed \cite{mcbride1991resonance}. It is clear in the figure \ref{Fig:4} that when the sample was measured with the wavelength of 633 nm (1.95 eV) the peaks were better defined and new peaks appeared at different frequencies which could be the  resonance effects. No peaks related to vibrations of compounds containing carbon were detected. The absence of those  peaks is indicative of  the purity of the samples.

\begin{table}
\centering
\caption{Raman peaks obtained from Fig. \ref{Fig:3}.}
\label{tab:1}       
\begin{tabular}{|p{1.3cm}|p{1.3cm}|p{1.3cm}|}
\hline\noalign{\smallskip}
2.33 eV&1.95 eV& 1.59 eV\\
\noalign{\smallskip}\hline\noalign{\smallskip}
273.01&277.73&278.24 \\
425.09&331.93&401.66 \\
620.62&409.36&519.3 \\
680.02&552.02&630.83 \\
1062.13&628.80&661.04 \\
1257.21&695.28&911.74 \\
&1057.43& \\
&1142.61& \\
&1254.29& \\
\noalign{\smallskip}\hline
\end{tabular}
\end{table}

\begin{figure}[h!] 
\centering
\includegraphics[width=0.5\textwidth]{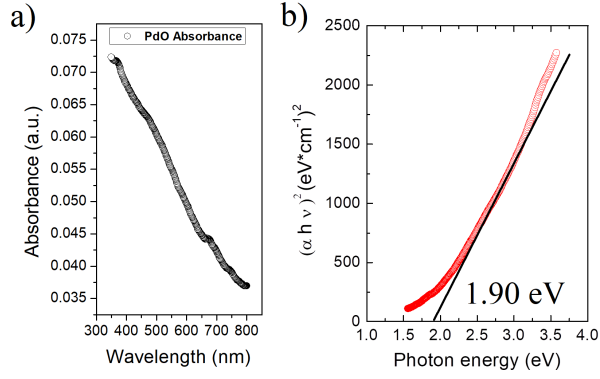} 
\caption{ \textbf{(a)} Optical absorption of the PdO nanoparticles, \textbf{(b)} Tauc plot for direct transition. The bandgap was estimated as 1.9 eV}
\label{Fig:5}       
\end{figure}

The optical properties of the single crystal particles of PdO were measured. Figure \ref{Fig:5} (a) shows the optical absorption spectra of the sample perform at room temperature. In this spectrum, we observe a strong absorption intensity in the near UV and violet region. From this measurement, the optical band gap was estimated from the absorbance spectrum using the Tauc$-$plot method \cite{li2006facile,kumar2020structural,ascencio2020comparative}. Considering a direct transition, the bandgap value was determined to be 1.90 eV as presented in the figure \ref{Fig:5} (b). The estimated bandgap values are in agreement with values proposed by \cite{huang2010growth} for PdO thin films, also in theoretical studies by DFT, close values have been obtained \cite{bruska2011electronic}. In different studies of PdO, some values have been found, which are in the range of [0.8$-$2.67] eV. The discrepancy of the values could be related to the structural defects at the surface, such as oxygen vacancies and confinement effects related to particle size reduction.

\subsection{Magnetism}
\label{Mag} 

\begin{figure}[h!] 
\centering
\includegraphics[width=0.52\textwidth]{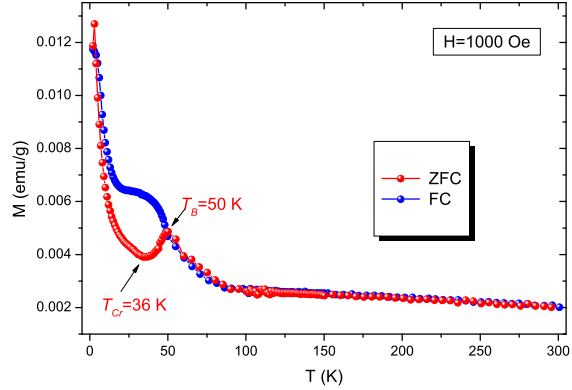} 
\caption{  Magnetization of PdO nanoparticles as  function of temperature  measured at 1000 Oe in ZFC and FC modes. The arrows marks the blocking temperature ($T_{B}$) and the crossover temperature ($T_{Cr}$)}
\label{Fig:6}       
\end{figure}

The magnetization M(T) as  function of temperature for PdO nanoparticles was determined with an applied magnetic field of 1000 Oe in two modes of magnetization, as usual;   zero field cooling (ZFC) and field cooling (FC) modes as  can be see in the figure \ref{Fig:6}. The difference between ZFC and FC curves indicates a large anisotopic field.

\begin{figure}[h!] 
\centering
\includegraphics[width=0.5\textwidth]{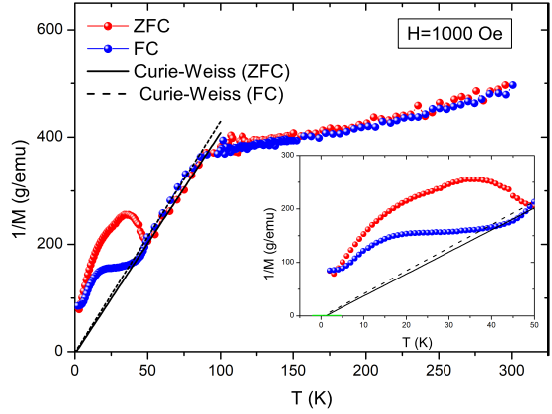} 
\caption{ A paramagnetic behavior was determined in the temperature range 100 - 300 of temperature with  the inverse of M as function of temperature. For temperatures between 50-100 Curie-Weiss fits gives a small $\theta_{C-W}$. Inset show the low temperature-depend of 1/M in order to appreciate the range of values for $\theta_{C-W}$}
\label{Fig:7}       
\end{figure}

In additionl we performed measurements of the inverse  magnetization versus temperature which  gives additional information about the behavior shown  figure \ref{Fig:7}. A clear paramgnetic behavior is seen from room temperature to about 100 K. Below  this temperature an incipient  ferromagnetism araises. The para-ferromagnetic transition has been observed in other nanoparticles and the transition was associated with oxygen vacancies \cite{sarkar2016paramagnetic}. At  low temperature the Curie Weiss temperature ($\theta_{C-W}$) behavior is very small, fitting gives values  between  -1.9 K or to about 4.5 K. However,  additional structure at  around 50 K, is seen in figure \ref{Fig:6}. This characteristic we related to  the metastable  states  when the nanoparticles are transported from one metastable state to the  other,  dependding of the applied  magnetic field. We related this characteristic or behavior with a macroscopic quantum tunneling \cite{xu2016quantum,tejada1996quantum}.  Two features give information of the magnetization; the blocking temperature $T_{B}$ \cite{bruvera2015determination} and the crossover temperature, $T_{Cr}$. The Fig. \ref{Fig:6} shows those characteristics temperatures at 50 K and 36 K respectively. Knowing $T_{B}=36$ K and the mean volume of the nanoparticles  ($V=6.54\SI{1e-20} cm^{3}$). The magnetic anisotropy constant ($K$) can be calculated by the the formula 1 \cite{li2010synthesis}, were $k_{B}$ is the Boltzman constant. 

\begin{equation}
K=25k_{B}T_{B}V^{-1}  
\end{equation}

The  value for $K$ is 1.89\SI{1e6} erg/cm$^{3}$ is not comparable with bulk PdO because this last is not magnetic.

Below $T_{B}$ the magnetization decrease with the temperature until 36 K, this temperature is the crossover temperature and separate the thermally activated regime $T_{Cr}$  from the quantum tunneling regime  \cite{zhang1996magnetic}. Below $T_{Cr}$ = 36 K, MQT effects take place and one can re-observed superparamgnetic state even when the thermal energy is smaller than the magnetic energy barrier \cite{xiao2012quantum,xu2016quantum}. 

\begin{figure}[h!] 
\centering
\includegraphics[width=0.5\textwidth]{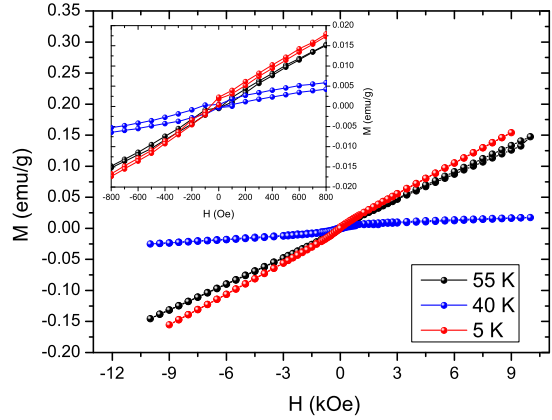} 
\caption{Magnetization of PdO nanoparticles as  function of H  at different temperatures. Inset shows the low field-dependent magnetization of the  nanoparticles. Hysteresis loop shows small hysteresis at 40 K in the blocked state }
\label{Fig:8}       
\end{figure}   
 
The magnetization of the nanoparticles was measure as a function of magnetic field H in order to confirm the superparamagnetism state below $T_{Cr}$ and above $T_{B}$. The field-dependent of magnetization was measured using three different temperatures, see figure \ref{Fig:8}. At the lowest temperature, 5 K,  there is no hysteresis, and the large magnetization values are related to superparamagnetic characteristics. This magnetic behavior corresponds with the expected characteristic in a quantum tunneling regime ($T<T_{Crossover}$). The value of the magnetization in the blocked state (40K) is less than the value of the magnetization measured at 5 K. Measurements present small hysteresis and a remanent magnetization M$_{r}$=\num{6.57e-4} emu/g with non symmetric coercive field   H$_{C+}$=82 Oe and H$_{C-}$=-128 Oe related with the possible coexistence antiferromagnetic state. Above the blocking temperature, 55 K, the behavior is superparamagnet, and the hysteresis is negligible. We assume that superparamagnetism is due to the small size of the particle ($\sim 5$ nm).

\section{Conclusions}
\label{Con}

In summary, this work presented a magnetic study of PdO nanoparticles in which we observed macroscopic quantum tunneling phenomena due to very small nanoparticles. The particle size was estimated at about 5.07 $\pm$ 1.97 nm from the TEM measurements. The Samples were analyzed by transmission electron microscopy in which the irregular shapes of the nanoparticles were observed. Using UV-VIS and Raman spectroscopy, the sample was analyzed spectroscopically. For the Raman studies, three excitation energies were used: 2.33, 1.95, and 1.59 eV ($\lambda$= 532, 633 and 780 nm), and observed the main peaks of PdO, besides the resonance phenomena, were observed. The UV-VIS measurements were performed, the behavior shows a strong absorption near to UV region. The bandgap value was estimated as 1.90 eV using Tauc-plot method. 
Finally, the most important aspect of this study is that we found a process related to macroscopic quantum tunneling of magnetization at low temperatures.

\section{acknowledgements}
Thanks to M.Sc. A. Bobadilla for the He supply. To R. Hernandez for the technical assistance in TEM. To A. Pompa and A. Lopez for technical support. We acknowledge DGAPA-PAPIIT Grant No. IT101920, F. Ascencio thanks to the DGAPA-UNAM, for the support through the Postdoctoral Scholarship

\bibliographystyle{spphys}      

\bibliography{Macroscopic_quantum_tunneling_in_PdO_nanoparticles}   

\end{document}